\documentclass[twocolumn,aps,prb,superscriptaddress,nofootinbib,float,longbibliography]{revtex4-2}
\usepackage[colorlinks,bookmarks=false,citecolor=blue,linkcolor=red,urlcolor=blue]{hyperref}
\usepackage{verbatim}
\usepackage{amsmath,amssymb,bm,braket,float,mathtools}
\usepackage{graphicx,xfrac,appendix}
\usepackage[english]{babel}
\makeatletter
\usepackage{epstopdf}
\makeatother

\begin{document}
\title{Long-range hopping in the quasi-periodic potential  weakens the non-Hermitian skin effect}
\author{Dechi Peng}
\affiliation{Institute of Quantum Precision Measurement, State Key Laboratory of Radio Frequency Heterogeneous Integration, College of Physics and Optoelectronic Engineering, Shenzhen University, Shenzhen 518060, China}
\affiliation{Department of Physics, Zhejiang Normal University, Jinhua 321004, China}
\author{Shujie Cheng}
\affiliation{Xingzhi College, Zhejiang Normal University,  Lanxi 321100, China}
\author{Gao Xianlong}
\thanks{gaoxl@zjnu.edu.cn}
\affiliation{Department of Physics, Zhejiang Normal University, Jinhua 321004, China}
\date{\today}

\begin{abstract}
In this paper, we investigate a non-Hermitian Aubry-André-Harper  model characterized by power-law hoppings ($1/s^{a}$) and a quasi-periodic parameter $\beta$,
where $a$ denotes the power-law index,  $s$ represents the hopping distance, and  $\beta$ belongs to the metallic mean family. In the intermediate phases, we find that ergodic states correspond to complex eigenvalues, multifractal states correspond to real eigenvalues, and localized states may exhibit either complex or real eigenvalues. 
In the localized phase, the energy spectra are complex, with most of the eigenvalues of localized states being real, while a few are complex.
Under open boundary conditions, our analysis of fractal dimensions and eigenstates reveals that all ergodic states transform into scale-free localized states. Furthermore, we demonstrate that long-range hoppings weaken the skin effect, offering another perspective for exploring non-Hermitian skin effects.

\end{abstract}
\maketitle

\section{introduction}

Localization phenomena in physics have attracted considerable interest in both single-particle and many-body systems. Notable examples include Anderson localization in single-particle systems \cite{PhysRev.109.1492}, many-body localization\cite{PhysRevB.107.035129,Bordia_2017,PhysRevLett.123.090603,RevModPhys.91.021001,PhysRevX.7.021013,PhysRevLett.126.080602}, and scar states \cite{PhysRevB.98.155134,PhysRevB.98.235156,PhysRevLett.125.230602} in many-body contexts.  
The study of Anderson localization demonstrates through scaling theory \cite{PhysRevLett.42.673,PhysRevLett.100.013906,PhysRevLett.105.163905} that  there is no ergodic phase in the systems under arbitrarily weak disorder in one (1D) and two (2D) dimensional models. However, in three dimensional (3D) scenarios, a mobility edge emerges within the single-particle spectra, delineating ergodic from localized states.
Beyond traditional Anderson models, mobility edges manifest in a class of generalized Aubry-André-Harper (AAH) models. It is well-known that the standard AAH model lacks a mobility edge \cite{aubry1980analyticity,1982JETP...56..612S,wilkinson1984critical}, only when the self-duality is broken by introducing, such as, 
the next-nearest-neighbor hoppings \cite{PhysRevB.83.075105}, the exponentially long-range hoppings \cite{PhysRevLett.104.070601,PhysRevB.103.134208}, the off-diagonal incommensurate hoppings \cite{PhysRevLett.127.116801,PhysRevB.106.144208,PhysRevB.104.085401,PhysRevB.103.134208,PhysRevB.103.014203}, the power-law hoppings \cite{PhysRevLett.123.025301,PhysRevB.103.075124,PhysRevB.83.075105,PhysRevB.100.174201}, the slow-varying potentials \cite{Cheng_2022,PhysRevB.105.104201}, and the generalized incommensurate potentials\cite{Farchioni1993IncommensuratePA,PhysRevB.101.064203,YUCE20142024,PhysRevA.95.062118,PhysRevB.100.054301,PhysRevLett.122.237601,PhysRevResearch.2.033052,PhysRevB.103.014203}.
Studies on single-particle mobility edge\cite{PhysRevLett.120.160404,PhysRevA.94.033615,PhysRevA.98.013635,PhysRevB.90.054303,Roy_2021,PhysRevB.101.064203,PhysRevB.96.085119,PhysRevB.103.144202,PhysRevB.102.024205,PhysRevA.105.063327,PhysRevB.105.205402} have greatly enhanced our understanding of how mobility edges influence thermalization and many-body
localization in interacting quasidisordered extensions. 

Meanwhile, non-Hermitian physics has garnered significant attention and has been extensively explored in quasi-periodic systems\cite{PhysRevB.103.134208,PhysRevB.106.144208,PhysRevB.102.024205,PhysRevB.104.085401,PhysRevB.103.014203,PhysRevB.107.174205,PhysRevB.104.224204,PhysRevLett.121.086803,LIAO2024107372,10.21468/SciPostPhys.16.1.029,PhysRevLett.132.063804,PhysRevLett.132.113802,PhysRevLett.129.070401,PhysRevB.100.054301,PhysRevA.105.063327,PhysRevA.103.033325,PhysRevB.101.174205,manna2023inner}. This exploration includes phenomena such as the non-Hermitian skin effect, the relationship between the real-complex transition of energy spectra and the localization transition of eigenstates, as well as the topological properties of non-Hermitian spectra\cite{PhysRevB.106.144208,PhysRevB.102.024205,PhysRevB.104.085401,PhysRevB.103.014203,PhysRevB.107.174205,PhysRevB.104.224204,PhysRevB.100.054301,PhysRevA.105.063327,PhysRevA.103.033325,PhysRevB.101.174205}. Among these topics, investigations into the non-Hermitian skin effect have been particularly comprehensive, encompassing high dimensional skin effects, size-dependent skin effects, and their associated topological characteristics\cite{PhysRevB.110.014106,PhysRevB.103.134208,PhysRevB.103.014203,PhysRevB.100.054301}. Despite these advancements, there remains a lack of studies exploring physical quantities influencing the degree of the non-Hermitian skin effect, beyond merely adjusting the non-Hermitian strength\cite{PhysRevB.103.195414}. In the present study, we aim to identify a significant physical quantity that impacts the degree of the non-Hermitian skin effect.

The relationship between the real-complex transition of eigenvalues and the localization transition of eigenstates has uncovered several intriguing phenomena. Notably, it has been observed that eigenvalues tend to be real in the localized phase, while exhibiting complex values in the ergodic phase\cite{weidemann2022topological,PhysRevB.100.054301}. However, there are instances where eigenvalues are complex in the localized phase and real in the ergodic phase \cite{PhysRevLett.122.237601,PhysRevB.104.224204,PhysRevB.107.174205,PhysRevB.101.174205}. Additionally, both real and complex eigenvalues coexist in the ergodic phase\cite{PhysRevA.103.033325}. Building upon these findings, this study seeks to explore another relationship that further elucidates the connection between eigenvalue behavior and state localization.

Recently, there has been growing interests in studying mobility edges in a class of generalized AAH model featuring power-law hoppings \cite{PhysRevLett.123.025301,PhysRevB.83.075105,PhysRevB.103.075124,PhysRevB.104.224204,PhysRevB.107.174205,PhysRevA.110.012222}, which can be induced by power-law interactions \cite{PhysRevB.83.075105,PhysRevLett.120.160404}. Deng et al. found that
when the power-law index $a<1$, there exist ergodic-to-multifractal (EM) edges in the intermediate regimes, while $a>1$ leads to ergodic-to-localized (EL) edges\cite{PhysRevLett.123.025301}.
Liu et al. examined a non-Hermitian AAH model incorporating nonreciprocal power-law hoppings and found that each ergodic state transforms into  a kind of special skin-like  state, i.e., scale-free localized (SFL) state under open boundary conditions (OBCs). Consequently, EM edges become skin-to-multifractal edges, and EL edges transition to skin-to-localized edges \cite{PhysRevA.110.012222}. 
However, our findings indicates that the fractal dimension of  the SFL states characterized by  long-range hoppings ($a<1$) significantly differs from those with short-range hoppings ($a>1$), indicating a variance in the localization degree of  the SFL states. This observation opens avenues for studying the degree of non-Hermitian skin effect.
In addition, in Liu's study\cite{PhysRevA.110.012222}, it is observed that within the intermediate phases where $a>$1, the eigenvalues corresponding to localized states include a few complex numbers alongside real ones. This has inspired us to examine the spectral properties of the localized phase within the same system.

The paper is organized as follows. In Sec. \ref{S2}, we present the Hamiltonian of the
non-Hermitian AA model with nonreciprocal power-law hopping and introduce the metallic mean family.
In Sec. \ref{S3}, for a fixed non-Hermitian strength, we study localization properties and spectral properties under both periodic boundary conditions (PBCs) and OBCs. In Sec. \ref{S4}, we study the effects of long-range hoppings on the skin effect of eigenstates, again considering a fixed non-Hermitian strength and OBCs. Finally, we provide a summary of our findings in Sec. \ref{S5}.

\section{MODEL AND HAMILTONIAN}\label{S2}
We consider a 1D non-Hermitian AAH model with nonreciprocal  power-law hoppings, expressed as:	
\begin{equation}\label{eq1}
  H =-J \sum_{j} \sum_{s=1}^{\infty} \frac{1}{s^{a}}\left({e^{k}c_{j}^{\dagger} }c_{j+s}+{{e^{-k}c_{j+s}^{\dagger} }c_{j} }\right) 
  + \sum_{j} \Delta_{j} c_{j}^{\dagger} c_{j},
  \end{equation}
where $J$ is set as the unit of energy. The power-law index $a$ governs the hopping distance of single particle in the system, with increasing $a $ leading to decreasing hopping distance. In the limit $a\rightarrow \infty$, this non-Hermitian system reverts to the nearest-neighbor hopping scenario \cite{PhysRevB.100.054301}. $J/s^{a}$ represents the power-law hopping strength between site $j$ and site $j+s$, while $e ^ {k} $ and $e ^ {- k} $ modulate nonreciprocal left and right hoppings, respectively. The non-Hermitian effect is introduced by the natural index $k$, which is fixed at $k=0.5$ for this study. When $k=0$, the model returns to its Hermitian form \cite{PhysRevLett.123.025301,PhysRevB.103.075124}, where EM edges are uncovered. 
The on-site potential $\Delta_{j}= \Delta \cos (2 \pi\beta j)$ is defined using $\beta$, which corresponds to the metallic mean family, derived from a generalized $\mu$-Fibonacci
recurrence relation: $F_{v+1}=\mu F_v+F_{v-1}$, with $F_0=0$ and $F_1=1$. The golden mean, $\beta=\beta_{g}$ is obtained by the limit
$\beta_{g}=\lim _{v \rightarrow \infty} F_{v-1} / F_v$ for $\mu=1$. Additionally, this recurrence can yield another metallic mean,
such as the silver mean $\beta=\beta_s=\sqrt{2}-1$  when $\mu=2$, and the bronze mean $\beta=\beta_b=(\sqrt{13}-3)/2$ when $\mu=3$. As a representative case, we focus on $\beta=\beta_{g}$ in this study.

\section{ Localization properties and spectral properties under periodic and open boundary conditions}\label{S3}

\begin{figure}[t]
	\centering
	\includegraphics[width=0.5\textwidth]{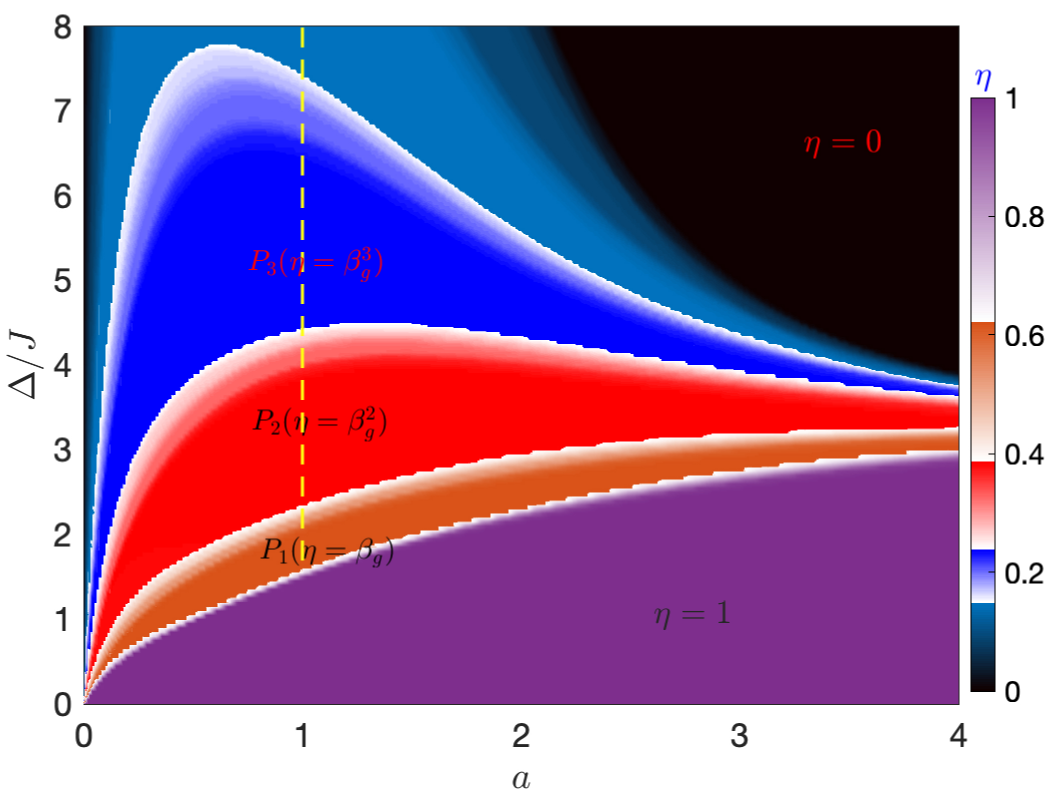}
	\renewcommand\figurename{Fig}
	
	
\renewcommand\figurename{Fig}
\caption{Phase diagram of the non-Hermitian AAH model as a function of the power-law hopping index $a$ and the potential strength $\Delta$, with $\beta_g =610/987$, non-Hermitian strength $k=0.5$, and system size $L = 987$ under PBCs. The colorbar represents the fraction of ergodic eigenstates $\eta $.  The diagram reveals three distinct phases: the ergodic phase (purple region), characterized by $\eta=1$; the localized phase (black region), characterized by $\eta=0$; and an intermediate phase where $0<\eta<1$. Within this intermediate phase, distinct regimes $P_{\ell=1,2,3,...}$ are identified, where $\eta=\beta^{\ell}_g$. For instance, $\eta=\beta_g$ (orange region, marked as $P_{1}$), $\eta=\beta^{2}_g$ (red region, marked as $P_{2}$), and $\eta=\beta^{3}_g$ (blue region, marked as $P_{3}$). The vertical yellow dashed line separates the long-range hopping regime ($a<1$) from the short-range hopping regime ($a>1$). White patches represent the normal intermediate phase, where the fraction of ergodic eigenstates $\eta$ does not depend on the quasiperiodic parameter $\beta$. These white patches can also be interpreted as transition intervals between two adjacent $P_{\ell}$ regimes.}\label{f1}
\end{figure}

 To investigate the relationship between the localization transition of eigenstates and the real-complex transition of eigenvalues under both PBCs and OBCs, we set the non-Hermitian strength at $k=0.5$ and the quasi-periodic parameter to $\beta=\beta_{g}$.
The phase diagram of the model under PBCs, as specified in Eq. (\ref{f1}), is illustrated in Fig.~\ref{f1}. The colorbar denotes the fraction of ergodic eigenstates $\eta$, defined as $\eta=\frac{N_e}{N_t}$, where $N_e$ represents the number of ergodic states and $N_t$ is the total number of eigenstates. 
We observe that the non-Hermitian system preserves features akin to those found in its Hermitian counterpart\cite{PhysRevLett.123.025301}. 
As depicted in the phase diagram, besides the ergodic (purple region) and localized (black region) phases---characterized by fractions of ergodic eigenstates $\eta = 1$ and $\eta = 0$, respectively---there exists an intermediate phase where $0<\eta<1 $. In addition, the vertical yellow dashed line separates the long-range hopping regime ($a<1$) from the short-range hopping regime ($a>1$) within intermediate phase.
In particular, within this intermediate phase, there are distinct $P_{\ell}$ regimes where the lowest $\beta^{\ell}_{g}$L eigenstates
are ergodic, exhibiting fractions  $\eta=\beta_g$ (orange region, marked by $P_{1}$),
$\beta_g^2$ (red region, marked by $P_{2}$), $\beta_g^3$ (blue region, marked by $P_{3}$). The fraction of ergodic eigenstates $\eta$ exhibits different dependencies in the $P_{\ell}$ regimes due to variations in the metallic mean. For example, if we take the silver mean $\beta=\beta_s=\sqrt{2}-1$ as the quasi-periodic parameter, the distribution pattern of $\eta$ changes, following $\eta=\beta_{s}+\beta^{2}_{s}$ $(P_{1})$, $\beta_{s}$ $(P_{2})$, $\beta^{2}_{s}+\beta^{3}_{s}$ $(P_{3})$, similar to the Hermitian cases\cite{PhysRevB.103.075124}.
In contrast to the Hermitian cases \cite{PhysRevLett.123.025301}, a notable distinction lies in how the $P_{\ell}$ regimes are analogous but separated by normal intermediate regimes (white patches) under the non-Hermitian influence, where the fraction of ergodic states is $\beta$-independent. This indicates that as $\Delta$ increases, the fractions of ergodic eigenstates $\eta$ between every two adjacent $P_{\ell}$ regimes do not exhibit a sudden shift, as observed in the Hermitian case\cite{PhysRevLett.123.025301,PhysRevB.103.075124}. Instead, they display a transition interval. An increase in $k$, corresponding to a stronger non-Hermitian effect, leads to a broader transition interval, which manifests as a larger expanse of white patches between two adjacent $P_\ell$ regimes. Theoretically, there are infinite $P_{\ell}$ regimes beyond the three illustrated in this phase diagram, as the potential strength $\Delta$  increases. Meanwhile, the effect of non-Hermitian expands the parameter range of the ergodic phase, while reducing that of the localized phase.

 Our primary conclusions are outlined as follows:
 \begin{enumerate}
    \item Under PBCs, as shown in the phase diagram in Fig.~\ref{f1}, $P_{\ell}$ regimes exist where the fraction of ergodic states follows $\eta= \beta^{\ell}_g$, similar to the Hermitian case. Additionally, long-range ($a<1$) and short-range ($a>1$) hoppings scenarios are characterized by EM and EL edges, respectively.
    \item In the intermediate phase under PBCs, eigenvalues associated with multifractal states are real, while those corresponding to localized states can be either complex or real. In the localized phase (black region), a significant portion of eigenvalues are complex, with a smaller fraction being real. This behavior aligns with the real-complex characteristics observed in the intermediate phase for localized states.
    \item Under OBCs, each ergodic state transforms into an SFL state. This transformation shifts EM edges to skin-to-multifractal edges and EL edges to skin-to-localized edges. Furthermore, as the power-law index $a$ increases, the localization degree of SFL states becomes more and more pronounced, indicating the weakening of the skin effect due to long-range hopping.
    \item Under OBCs, the eigenvalues corresponding to SFL states exhibit complexity, while maintaining similar real-complex energy spectral properties for multifractal states and localized states as observed under PBCs.
 \end{enumerate}

\begin{figure}[htp]
		\centering
		\includegraphics[width=0.5\textwidth]{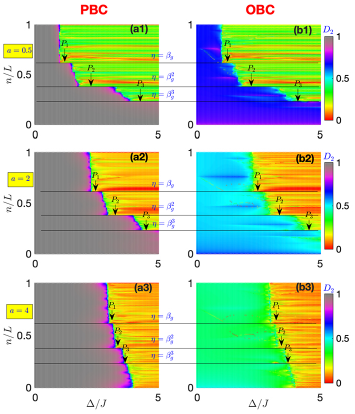}
		\renewcommand\figurename{Fig}
	\caption{  Fractal dimension $D_2 $ (shown in color) of different eigenstates as a function of potential strength $\Delta$, with $k=0.5$, system size $L=987$, box size $d=3$,
	and golden mean $\beta_g =610/987$. Results are shown for (a1) $a=0.5$, (a2) $a=2$, and (a3) $a=4$ under PBCs, and for (b1) $a=0.5$, (b2) $a=2$, and (b3) $a=4$ under OBCs. Eigenstates are arranged in ascending order based on the real part of the eigenvalues, indexed by  $n$.}\label{f2}
\end{figure}
To further investigate the relationship between the localization properties of eigenstates and the real-complex transformation of eigenvalues, we utilize the fractal dimension to quantitatively assess the degree of localization of eigenstates.
The fractal dimension $D_{f}$ is defined based on the box counting procedure \cite{PhysRevLett.62.1327,doi:10.1142/S021797929400049X,RevModPhys.67.357,PhysRevB.68.184206}
and is given by:
\begin{equation}\label{eq2}
D_f=\lim _{L_{d} \rightarrow \infty } \frac{1}{1-f} \frac{\ln \sum_{m=1}^{L_{d}}\left(\mathcal{I}_m\right)^f}{\ln L_{d}},
\end{equation}
where $L_d=L / d$ is the number of the box, with $L$ being the system size and $d$ the box size. The parameter $f$ is the scale index, and $\mathcal{I}_m=\sum_{j \in m}\left|u^{n}_j\right|^2$ corresponds to the probability of
detecting within the $m$$th$ box for the $n$$th$ normalized eigenstate $|u^{n}_j\rangle $.
Without loss of generality, we focus on the fractal dimension $D_{2}$. Under PBCs, with a system size of $L=987$ and a box size of $d=3$, alongside the golden mean  $\beta_{g}=610/987$, we present the plot of $D_{2}$ for all eigenstates as a function of potential strength $\Delta$. For $a=0.5$ (long-range hopping), the results are shown in Fig.~\ref{f2}(a1), while for $a=2.0$ and $a=4.0$ (short-range hopping), the results are depicted in Figs.~\ref{f2}(a2) and \ref{f2}(a3), respectively. In the $P_{\ell}$ regimes, two types of edges exhibit a step-wise dependence on $\Delta$, corresponding to $\eta=\beta^{\ell}_{g}$.
Under OBCs, using the same parameters as in the PBCs case, we plot $D_{2}$ for all eigenstates as a function of $\Delta$. For $a=0.5$ (long-range hopping), the results are shown in  Fig.~\ref{f2}(b1), while for $a=2.0$ and $a=4.0$ (short-range hopping), the results are presented in Figs.~\ref{f2}(b2) and \ref{f2}(b3), respectively. Consistent with the observations under PBCs, the $P_{\ell}$ regimes under OBCs also exhibit a step-wise dependence on $\Delta$.
Upon comparing the fractal dimension $D_{2}$ values, two phenomena emerge that distinguish OBCs from PBCs. Firstly, under the same power-law hopping index $a$, the fractal dimensions $D_{2}$ below each step differ from those in the PBCs case. This difference arises due to the presence of SFL states below the steps, instead of ergodic states as observed under PBCs. Secondly, as the power-law hopping index $a$ increases, the fractal dimension $D_{2}$ below each step decreases under OBCs, in contrast to the behavior under PBCs. This decrease is attributed to the enhanced localization degree of the SFL states with increasing $a$. Additional insight into these phenomena will be provided below.

\begin{figure}[htp]
	\centering
	\includegraphics[width=0.5\textwidth]{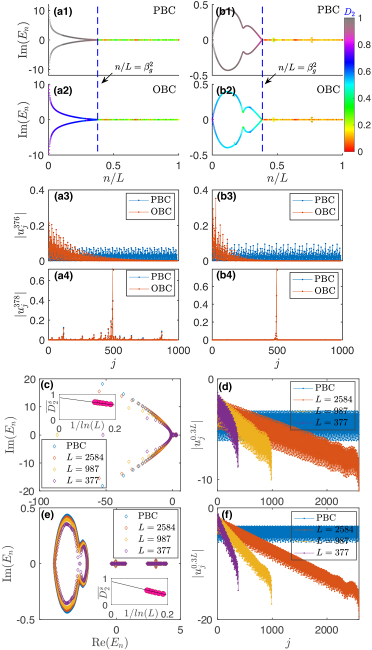}
	\renewcommand\figurename{Fig}
	\caption{(a1)-(a2) Imaginary part of the energy spectra Im $(E_n )$ under PBCs and OBCs, respectively, sorted in ascending order of the real part. The colorbar represents the fractal dimension $D_{2} $. Parameters are $L=987 $, $\beta_{g}=610/987 $, $d=3 $, $k=0.5 $, $a=0.5 $, $\Delta/J=2.6 $. (a3)-(a4) Probability amplitude of the eigenstate corresponding to the 376$th$ and 378$th$ eigenvalue under PBCs and OBCs, respectively, with $a=0.5 $, $\Delta/J=2.6 $.  (b1)-(b2) Imaginary part of the energy spectra, Im $(E_n )$, for $a=2.0 $ and $\Delta/J=3.6 $ under PBCs and OBCs, respectively, also sorted in ascending order of their real part. The colorbar represents the fractal dimension $D_{2} $. (b3)-(b4) Probability amplitude of the eigenstate corresponding to the 376$th$ and 378$th$ eigenvalue under PBCs and OBCs, respectively, with $a=2.0 $ and $\Delta/J=3.6 $. 
(c)-(e)  Energy spectra in the complex plane for different sizes with $a=0.5$, $\Delta/J = 2.6$, and $a=2.0$, $\Delta/J = 3.6$, respectively. The diamonds correspond to PBCs with $L = 2584$.
The insert figures of (c) and (e) display the average fractal dimension $\overline{D^{s}_2}$ versus L for SFL states with $a=0.5$, $\Delta/J = 2.6$, and $a=2.0$, $\Delta/J = 3.6$, respectively.
(d)-(f) Probability amplitude of the $0.3L$$th$ eigenstates (in log scale) for different system sizes with $a=0.5$, $\Delta/J = 2.6$, and $a=2.0$, $\Delta/J = 3.6$, respectively. PBCs are represented with $L = 2584$.}\label{f3}
\end{figure}
To further investigate the localization properties of eigenstates under both PBCs and OBCs, and to understand the relationship between the localization transition of eigenstates and the real-complex transition of eigenvalues, we plot the imaginary part of the energy spectra with fixed parameters. Additionally, we analyze the fractal dimension $D_{2}$  of the eigenstates corresponding to these eigenvalues, with eigenvalues sorted by the ascending order of their real parts.

Under PBCs, Figs.~\ref {f3}(a1) and \ref {f3}(b1) illustrate the imaginary part of the energy spectra and the fractal dimension $D_{2}$ of the corresponding eigenstates for the case of long-range hoppings ($a<1$) and short-range hoppings ($a>1$), respectively. The parameters in Figs.~\ref {f3}(a1) and \ref {f3}(b1) are selected from the $P_2 $ regime (The results for the $P_1 $ regime are presented in Appendix \ref{A}): specifically, $a=0.5 $, $\Delta/J=2.6 $ for long-range hoppings, and $a=2.0 $, $\Delta/J=3.6 $ for short-range hoppings. 
As shown in Fig.~\ref {f3}(a1), the eigenvalues undergo a complex-real transition at $n/L=\beta_g ^ 2 $. The fractal dimension analysis reveals that for $n/L<\beta_ {g} ^ 2 $, the fractal dimension $D_ {2} \rightarrow1 $, indicating that the eigenstates are ergodic. Conversely, for $n/L>\beta_ {g} ^ 2 $, $D_ {2} $ tends to a finite value, indicating that the eigenstate is multifractal, thus marking the appearance of an EM edge at $n/L=\beta_g ^ 2$.  To further elucidate the presence of EM edges, we examine the probability amplitudes of the eigenstates corresponding to the  $(n-1)th$ and the $(n+1)th$ eigenvalues at the energy index $n/L=\beta_g ^ 2 $. As shown in Figs.~\ref {f3}(a3) and \ref {f3}(a4), for $L=987 $, $n/L=\beta_g ^ 2=377/987 $, the eigenstate $u^ {376}_ {j} $ is ergodic, while the eigenstate $u^ {378}_ {j} $ is multifractal, confirming the existence of an EM edge at $n/L=\beta_g ^ 2 $. This finding also indicates that the eigenvalues corresponding to ergodic states are complex, whereas those linked to multifractal states are real. However, exceptions arises in non-Hermitian systems exhibiting Parity-Time ($PT$) symmetry, which involve power-law hoppings and complex potential\cite{PhysRevB.107.174205,PhysRevB.104.224204}. In these systems, the eigenvalues for ergodic states are real, while those for the multifractal states are complex.
As illustrated in Fig.~\ref{f3}(b1), the fractal dimension analysis shows that when $n/L<\beta_ {g} ^ 2 $, the fractal dimension of the eigenstate approaches $D_ {2}\rightarrow1$, indicating ergodic behavior. Conversely, when $n/L>\beta_ {g} ^ 2 $, $D_ {2}\rightarrow0$, signifying localization and the presence of EL edge at $n/L=\beta_g ^ 2 $. 
As shown in Figs.~\ref {f3}(b3) and \ref {f3}(b4), when $L=987 $, $n/L=\beta_g ^ 2=377/987 $, the eigenstate $u^ {376}_ {j} $ is ergodic, while the eigenstate $u^ {378}_ {j} $ is localized, confirming the existence of an EL edge at $n/L=\beta_g ^ 2 $. Additionally, is is observed that the eigenvalues for $n/L<\beta_g ^ 2 $ are complex, indicating that the eigenvalues of ergodic states are complex. However, for $n/L>\beta_g ^ 2 $, the eigenvalues can be either real or complex, suggesting that the eigenvalues corresponding to localized states may similarly exhibit real or complex characteristics. The spectral properties of the eigenvalues and the localization behavior of the eigenstates remain consistent with the above descriptions even at larger system sizes (see Appendix \ref{B}). 

Under OBCs, in Fig.~\ref{f3}(a2) with $a=0.5$, compared to Fig.~\ref{f3}(a1), a significant change in the fractal dimension $D_{2}$ is observed for $n/L<\beta_ {g}^2 $, indicating an alteration in the localization of the eigenstates. Figs.~\ref{f3}(a3) and \ref{f3}(a4) demonstrate that eigenstates with indices $n/L<\beta_ {g}^2 $ transform skin-like states, whereas those with $n/L>\beta_ {g}^2 $ remain multifractal.
In Fig.~\ref{f3}(b2) with $a=2.0$, compared to Fig.~\ref{f3}(b1), the fractal dimension $D_{2}$ undergoes a significant change for $n/L<\beta_{g}^2 $, reflecting changes in the localization of the eigenstates. Figs.~\ref{f3}(b3) and \ref{f3}(b4) reveal that the eigenstates with indices $n/L<\beta_{g}^2$ become skin-like states, while those with $n/L>\beta_{g}^2 $ remain localized. These findings are consistent with those reported in reference \cite{PhysRevA.110.012222}. Moreover, comparing Fig.~\ref{f3}(a2) with Fig.~\ref{f3}(b2), a significant difference in fractal dimension is evident for $n/L<\beta_{g}^2 $, indicating variations in the degree of localization of the skin-like state, which differs from the observation under PBCs. Furthermore, by comparing Figs.~\ref{f3}(a3) and \ref{f3}(b3), it is apparent that the degree of localization of the skin-like state differs between long-range $(a<1)$ and short-range $(a>1)$ hoppings. At the same time, compared to the case of PBCs,  we observe no significant change in the shape of the energy spectra, only in the magnitude of the values. The spectral properties of the eigenvalues and the localization behavior of the eigenstates are also consistent with the above descriptions at larger system sizes (see Appendix \ref{B}). 

It is important to note that the skin-like state discussed here does not correspond to a standard fully localized skin state, but rather represents the SFL state.
To demonstrate that the skin-like states correspond to SFL states, we present the eigenvalues in the complex plane for both $a>1$ and $a<1$ across varying system sizes [refer to Figs.~\ref{f3}(c) and \ref{f3}(e)]. It is observed that the eigenvalues in the region of $n/L<\beta_{g}^2$ exhibit a dependency on the system size, gradually approaching the behavior observed under PBCs. This phenomenon provides compelling evidence of SFL states, where eigenvalues under OBCs gradually converge towards those of PBCs as $L\rightarrow \infty$\cite{PhysRevA.110.012222}. In contrast, for $n/L>\beta_{g}^2$, the corresponding eigenvalues remain unaffected by system size and boundary conditions. The scale-free nature of the SFL states is further manifested in the eigenstates. Figs.~\ref{f3}(d) and \ref{f3}(f) depict the eigenstate corresponding to $n = 0.3L$ with $a=0.5$ and $a=2$, respectively. The probability amplitudes of these eigenstates are localized at the left boundary for smaller system sizes. However, as the system size increases, the localization length becomes preserved, leading to a gradual convergence of the probability amplitudes towards an extended state.
To further elucidate the extended characteristics of the SFL states, the average fractal dimension $\overline{D^{s}_2}$ of the SFL states is calculated as $\overline{D^{s}_2}=\sum_n \frac{1}{L_s} D_{2}^{s}(n)$, where $D_{2}^{s}$ is fractal dimension of the SFL state, and $L_s$ is the number of the SFL state. As shown in the inset figures of Figs.~\ref{f3}(c) and \ref{f3}(e), the average fractal dimension of the SFL states increases with growing $L$, providing substantial evidence of the extended properties of the SFL states\cite{PhysRevA.110.012222}.

\begin{figure}[htp]
	\centering
	\includegraphics[width=0.5\textwidth]{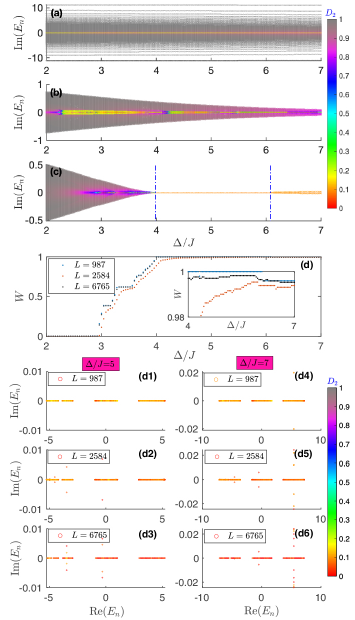}
	\renewcommand\figurename{Fig}
\renewcommand\figurename{Fig}
\caption{Variation of the imaginary part of the energy spectra, Im $(E_n)$, with the potential strength $\Delta $, for $k=0.5 $, $L=987 $, $\beta_{g}=610/987 $, and $d=3$. Panel (a) corresponds to the case of $a=0.5 $,  panel (b) corresponds to $a=2.0 $, and panel (c) corresponds to $a=4.0$.  The colorbar represents the fractal dimension  $D_2 $.  Panel (d) illustrates the variation of the fraction of real eigenvalues $W$ with respect to the potential strength $\Delta$ for $a=4.0$ across different sizes. For $a=4.0$ and $\Delta/J=5.0 $, panel (d1)-(d3) are the energy spectra for $L=987$, $L=2584$, and  $L=6765$, respectively. For $a=4.0$ and $\Delta/J=7.0 $, panel (d4)-(d6) show the energy spectra  for $L=987$, $L=2584$, and  $L=6765$, respectively. The colorbar represents the fractal dimension  $D_2 $. For $L=2584$, we use $\beta_{g}=1597/2584 $ and $d=4$. For $L=6765$, we use $\beta_{g}=4181/6765 $ and $d=5$. }\label{f4}
\end{figure}

To further demonstrate the universality relationship between the localization transition of eigenstates and the real-complex transition of eigenvalues in intermediate phases, we examine the variation of the imaginary part of the energy spectra Im $(E_n) $  with the potential strength $\Delta $ for both long-range ($a<1$) and short-range ($a>1$) hoppings. The localization of eigenstate is quantified using the fractal dimension $D_2 $. The similarities in the properties of the energy spectra under both PBCs and OBCs are evident, as shown in Figs.~\ref{f3}(a1), \ref{f3}(a2), \ref{f3}(b1) and \ref{f3}(b2). Therefore, we use PBCs as a representative case for further analysis.
As shown in Fig.~\ref{f4}(a), for long-range hoppings ($a=0.5$), within the mixed region where ergodic states and multifractal states coexist---multifractal states are confined to areas where Im $(E_n)=0$. This observation further supports that eigenvalues associated with ergodic states are complex, whereas those linked to multifractal states are real. Conversely, Fig.~\ref{f4}(b) illustrates that for short-range hoppings ($a=2.0 $), within the mixed region---where ergodic and localized states coexist--- localized states appear in regions where Im $(E_n)=0$ and Im $(E_n)\neq0$. Here, the eigenvalues corresponding to ergodic states are complex, while those for localized states may either be real or complex.

To investigate the real-complex transition of eigenvalues in the localized phase---where all eigenstates associated with the energy spectra are localized---we examine the variation of the imaginary part of the energy spectra Im $(E_n) $ as function of $\Delta $ for $a=4.0 $. The localization of the eigenstates is assessed using the fractal dimension $D_2 $.
As shown in Fig.~\ref{f4}(c), localized phases emerge as $\Delta $ increases. The imaginary part of the energy spectra Im $(E_n) $ becomes 0, although there exist ranges of $\Delta $ where Im $(E_n)\ne 0 $ at larger $\Delta $. This indicates that both real and complex energy spectra can exist in the localized phase, which is different from the scenarios where the energy spectra in the localized phase are entirely real\cite{weidemann2022topological,PhysRevB.100.054301} or entirely complex\cite{PhysRevLett.122.237601,PhysRevB.104.224204,PhysRevB.107.174205,PhysRevB.101.174205}.

Our investigation revealed that the presence of real energy spectra in the localized phase is attributed to finite-size effect, and at larger system sizes, the real energy spectra disappears. To elucidate this observation, we calculate the fraction of real eigenvalues, denoted as $W= \frac{N_r}{N_L}$, where $N_L$ represents the total number of eigenvalues, and $N_r$ denotes the count of real eigenvalues with an imaginary part satisfying $|Im(E)|<C$. Here, the cutoff value for the imaginary part is set as $C=10^{-13}$ to account for numerical diagonalization errors.
In Fig.~\ref{f4}(d), with a fixed value of $a=4.0$, we plot the variation of the fraction of real eigenvalues $W$ with the potential strength $\Delta$ for different system sizes. In the ergodic phase, it is observed that for all system sizes, $W=0$, indicating that all eigenvalues are complex. As the potential strength $\Delta$ surpasses $\Delta/J=4$, the system undergoes a transition into the localized phase. Specifically, for $L=987$, within a limited range of $\Delta$, $W=1$, indicating that all eigenvalues are real, a feature visually confirmed in Fig.~\ref{f4}(d1).
For higher values of $\Delta$, where $W<1$ but tends towards 1, this implies the existence of a small number of complex eigenvalues, as visually depicted in Fig.~\ref{f4}(d4), consistent with the observations in Fig.~\ref{f4}(c). However, at $L=2584$ and $L=6765$, the absence of $W=1$ suggests that, at larger system sizes, the scenario where all eigenvalues are real diminishes, with a small number of complex eigenvalues persisting, visually evident in Figs.~\ref{f4}(d2), \ref{f4}(d3), \ref{f4}(d5), and \ref{f4}(d6).
Furthermore, a detailed analysis of Figs.~\ref{f4}(d2), \ref{f4}(d3), \ref{f4}(d5), and \ref{f4}(d6) reveals that the maximum value of the imaginary part of complex eigenvalues remains constant as the system sizes increase, indicating the persistence of a small number of complex eigenvalues even as the system size $L$ approaches infinity.



\section{Impact of long-range hopping on skin effect}\label{S4}
In this section, we examine the impact of long-range hoppings on non-Hermitian skin effects under OBCs, with a fixed non-Hermitian strength $k=0.5$.
Our findings reveal that, under OBCs,  the degree of localization of these SFL states increases with the power-law index $a$, which governs the hopping distance. Specifically, shorter hopping distances correspond to stronger skin effects, and conversely, longer hopping distances result in weaker skin effects.

\begin{figure}[htp]
		\centering
		\includegraphics[width=0.5\textwidth]{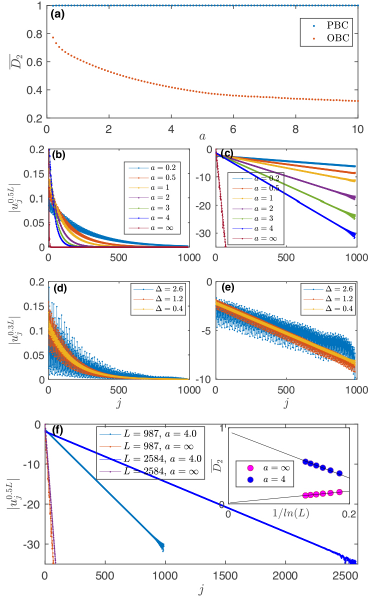}
		\renewcommand\figurename{Fig}
		
	\renewcommand\figurename{Fig}
	\caption{(a) Behavior of the average fractal dimension $\overline{D_2}$ as a function of the power-law index $a$ under PBCs and OBCs, with parameters $\Delta/J=0.4$, $k=0.5$,
	$\beta _g = 610/987$, and $L=987$. (b) Probability amplitude of the eigenstate corresponding to the $0.5Lth$ eigenvalue and (c) its distribution on a logarithmic scale for different power-law indices $a$, with  parameters $\Delta/J=0.4$, $k=0.5$, $\beta _g = 610/987$, and $L=987$. (d) Probability amplitude of the eigenstate corresponding to the $0.3L$$th$ eigenvalue and (e) its logarithmic distribution for different strengths of the potential $\Delta $, with parameters $a=0.5$,  $k=0.5$, $\beta _g = 610/987$, and $L=987$. (f) For a fixed value of $\Delta/J=0.4$, the investigation focuses on the behavior of the probability amplitude of the 0.5L$th$ eigenstate in a logarithmic scale across various system sizes for power-law indices $a=4$ and $a=\infty$, where the illustration shows that for $a=4$ and $a=\infty$, the average fractal dimension $\overline{D_2}$ changes with system sizes. }\label{f5}
	\end{figure}

To thoroughly investigate the impact of hopping distance on SFL state, we fix the potential strength $\Delta/J=0.4$ within ergodic phase (purple regime) as depicted in Fig.~\ref{f1} and  calculate the average fractal dimension $\overline{D_2}$, defined as $\overline{D_2}=\sum_n \frac{1}{L} D_{2}(n)$.
Considering the system size $L=987$, box size $d=3$, and with the golden mean $\beta_{g}=610/987$, we plot the average fractal dimension $\overline{D_{2}}$ of the full eigenstates
as a function of the power-law index $a$ for $\Delta/J=0.4$ under both OBCs and PBCs in Fig.~\ref{f5}(a). Under PBCs, the average fractal dimension $\overline{D_{2}}$ tends toward 1 and remains constant with increasing the power-law index $a$, indicating ergodic eigenstates whose localization does not vary with the hopping distance. However, under OBCs, an increase in $a$ causes $\overline{D_{2}}$ to decay from near 0.8, signifying that the localization of SFL states intensifies as the hopping distance decreases. 
These observations are further illustrated in the distribution of probability amplitudes of the eigenstates.

To clearly observe the influence of the power-law index $a$, which controls the hopping distance, on the non-Hermitian skin effect under OBCs, we present the probability amplitude distribution of the $ 0.5L$$th$ eigenstate for various values of $a$.
Considering a system size of $L=987$, potential strength $\Delta/J=0.4$,
and the golden mean $\beta_{g}=610/987$, we plot the probability amplitude and its logarithmic distribution in Figs.~\ref{f5}(b) and \ref{f5}(c). 
As depicted in Fig.~\ref{f5}(b), for both $a>1$ and $a<1$, the states exhibit characteristics of SFL states. With an increase in the power-law index $a$, the manifestation of the skin effect becomes more pronounced, leading to a heightened degree of localization in the SFL state. At $a=\infty$, the eigenstate becomes fully localized at the left boundary. To more clearly illustrate the varying degrees of localization of SFL states for different $a$, we plot the logarithm of the probability amplitude of the $0.5L$$th$ eigenstate. As shown in Fig.~\ref{f5}(c), the probability amplitude decays more rapidly with increasing $a$. This further illustrates the influence of power-law index $a$ on the skin effect, demonstrating that in this system, long-range hoppings weaken the skin effect. It is important to emphasize that at $a=\infty$, corresponding to nearest neighbor hopping, the skin-like states no longer exhibit the characteristics of SFL states and instead transform into fully localized states at the boundary\cite{PhysRevB.100.054301}.
To highlight the difference between finite values of $a$ and $a=\infty$, we plot the logarithm of the probability amplitude of the $0.5L$$th$ eigenstate for $a=4$ and $a=\infty$ across various system sizes. As depicted in Fig.~\ref{f5}(f), for $a=4$, the localization length remains preserved as the system size increases, a hallmark of the SFL state. Conversely, for $a=\infty$, the localization length is no longer maintained as the system size grows. Furthermore, the distinction between these two types of skin-like states can be further characterized by examining the average fractal dimension $\overline{D_2}$ as a function of the system size $L$. As illustrated in the inset of Fig.~\ref{f5}(f), for $a=4$, $\overline{D_2}$ increases with system size, indicating the presence of the SFL state. In contrast, for $a=\infty$, $\overline{D_2}$ decrease towards zero as the system size grows, characteristic of a fully localized state\cite{PhysRevA.110.012222,PhysRevB.100.054301}.

To assess whether the strength of the potential $\Delta $ impacts the non-Hermitian skin effect, particularly in light of the varying degrees of localization of SFL states illustrated in Fig.\ref{f3}, we focus on changes arguably induced by different power-law indices $a$ and potential strengths $\Delta $. By fixing the power-law index $a=0.5 $, system size $L=987$, and the golden mean $\beta_{g}=610/987$, we analyze probability amplitude distribution of the $0.3L$$th$ eigenstate and its logarithmic distribution across different  $\Delta $ values. As shown in Figs.~\ref{f5}(d) and \ref{f5}(e), variations in the potential strength do not significantly alter the degree of localization of the SFL state, even when viewed on a logarithmic scale. This suggests that the strength of the potential $\Delta $ does not affect the non-Hermitian skin effect. Thus, the differences in localization degrees in Fig.~\ref{f3} can be attributed solely to changes in the power-law index $a$. This phenomenon persists even in larger sizes (see Appendix \ref{C}).

\section{ conclusion}\label{S5}
In conclusion, we have investigated a non-Hermitian AAH model characterized by nonreciprocal power-law hoppings. We first presented the phase diagram of the non-Hermitian system, revealing that, compared to to the Hermitian case, the parameter range of the ergodic phase expands while that of the localized phase contracts. Additionally, the quasi-periodic parameter $\beta$-dependent $P_{\ell}$ regimes are delineated by normal intermediate regimes. We further elucidated the relationship between the real-complex transition of eigenvalues and the localization transition of eigenstates under PBCs. In the intermediate phases, ergodic states are associated with complex eigenvalues, multifractal states with real eigenvalues, and localized states may exhibit either complex or real eigenvalues.  Moreover, we revealed another relationship that there are complex energy spectra in both the localized phase and the ergodic phase. 
Most importantly, we demonstrate that long-range hoppings weaken the skin effect, thereby providing another perspective for the study of non-Hermitian skin effects.

\section{acknowledgements} 
Dechi Peng thanks Yongguan Ke for useful discussion. The authors acknowledge support from NSFC under
Grants No. 12174346. Shujie Cheng is supported by Zhejiang Provincial Natural Science Foundation of China under Grant No. LQN25A040012
\appendix
\section{ Localization properties and spectral properties in the $P_1$ regime}\label{A}

\begin{figure}[htp]
		\centering
		\includegraphics[width=0.5\textwidth]{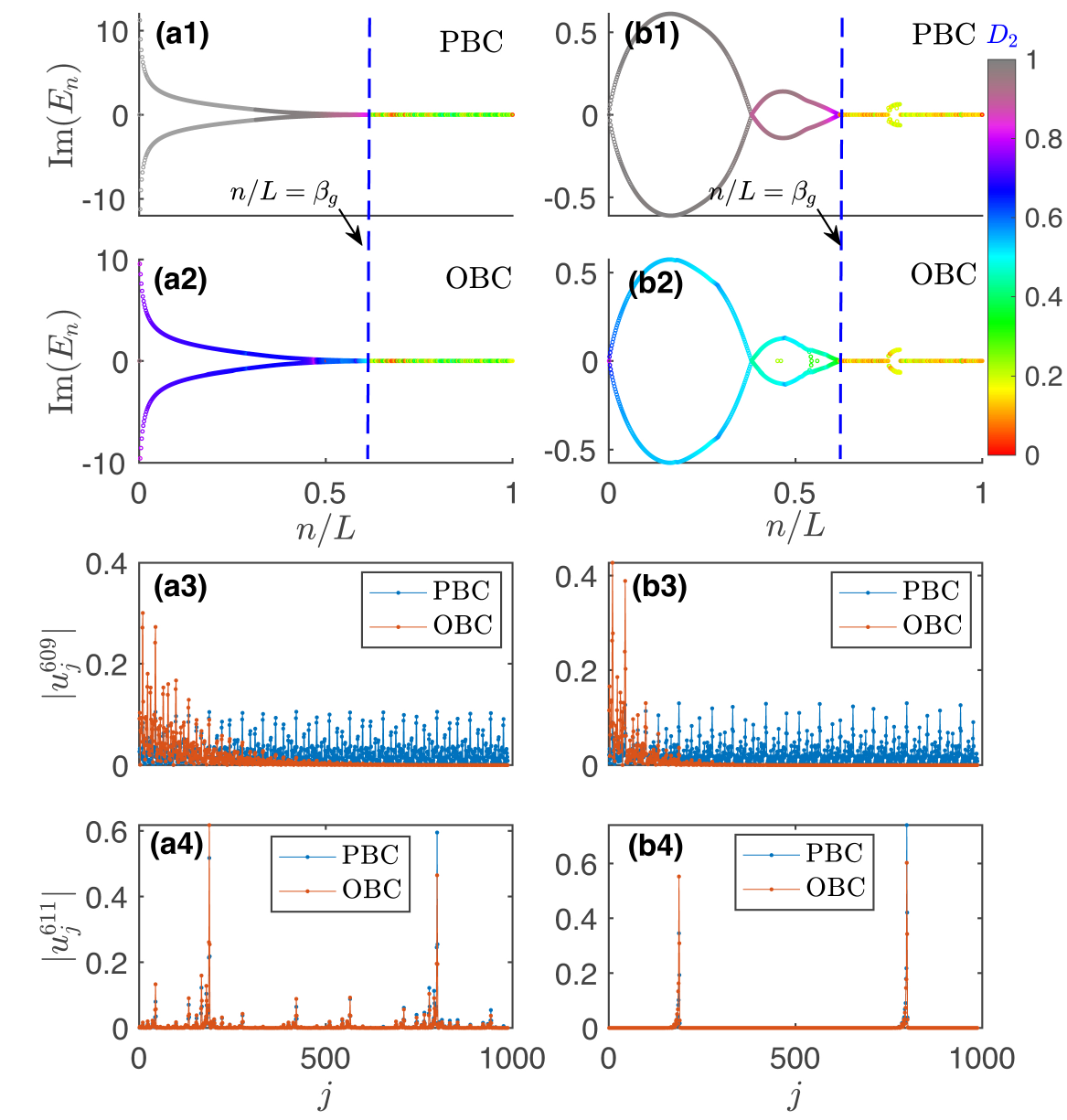}
		\renewcommand\figurename{Fig}
		\caption{
(a1)-(a2) Imaginary part of the energy spectra Im $(E_n )$ under PBCs and OBCs, respectively, sorted in ascending order of the real part. The colorbar represents the fractal dimension $D_{2} $. Parameters are $L=987 $, $\beta_{g}=610/987 $, $d=3 $, $k=0.5 $, $a=0.5 $, $\Delta/J=1.2 $. (a3)-(a4) Probability amplitude of the eigenstate corresponding to the 609$th$ and 611$th$ eigenvalue under PBCs and OBCs with $a=0.5 $, $\Delta/J=1.2 $.  (b1)-(b2) Imaginary part of the energy spectra Im $(E_n )$ for $a=2.0 $ and $\Delta/J=2.6 $ 
 under PBCs and OBCs, respectively, also sorted in ascending order of their real part, with the colorbar representing the fractal dimension $D_{2} $. (b3)-(b4) Probability amplitude of the eigenstate corresponding to the 609$th$ and 611$th$ eigenvalue under PBCs and OBCs with $a=2.0 $ and $\Delta/J=2.6 $.}\label{f6}
\end{figure}
To more fully illustrate the relationship between the $P_\ell$ regimes, we present the results for the $P_1$ regime.
Under PBCs, the parameters in Figs.~\ref {f6}(a1) and \ref {f6}(b1) are selected from the $P_1 $ regime 
specifically, $a=0.5 $, $\Delta/J=1.2$ for long-range hoppings, and $a=2.0 $, $\Delta/J=2.6 $ for short-range hoppings. 
As shown in Fig.~\ref {f6}(a1), the eigenvalues undergo a complex-real transition at $n/L=\beta_g$. The fractal dimension analysis reveals that for $n/L<\beta_ {g}  $, the fractal dimension $D_ {2} \rightarrow1 $, indicating that the eigenstates are ergodic. Conversely, for $n/L>\beta_ {g}  $, $D_ {2} $ tends to a finite value, indicating that the eigenstate is multifractal, thus marking the appearance of an EM edge at $n/L=\beta_g $.  To further elucidate the presence of EM edges, we examine the probability amplitudes of the eigenstates corresponding to the  $(n-1)th$ and the $(n+1)th$ eigenvalues at the energy index $n/L=\beta_g $. As shown in Figs.~\ref {f6}(a3) and \ref {f6}(a4), for $L=987 $, $n/L=\beta_g =610/987 $, the eigenstate $u^ {609}_ {j} $ is ergodic, while the eigenstate $u^ {611}_ {j} $ is multifractal, confirming the existence of an EM edge at $n/L=\beta_g  $. This finding also indicates that the eigenvalues corresponding to ergodic states are complex, whereas those linked to multifractal states are real.
As illustrated in Fig.~\ref {f6}(b1), the fractal dimension analysis shows that when $n/L<\beta_ {g}$, the fractal dimension of the eigenstate $D_ {2}\rightarrow1$, indicating ergodic behavior. Conversely, when $n/L>\beta_ {g} $, $D_ {2}\rightarrow0$, signifying localization and the presence of EL edge at $n/L=\beta_g$. 
As shown in Figs.~\ref {f6}(b3) and \ref {f6}(b4), when $L=987 $, $n/L=\beta_g =610/987 $, the eigenstate $u^ {609}_ {j} $ is ergodic, while the eigenstate $u^ {611}_ {j} $ is localized, confirming the existence of an EL edge at $n/L=\beta_g$. Additionally, is is observed that the eigenvalues for $n/L<\beta_g$ are complex, indicating that the eigenvalues of ergodic states are complex. However, for $n/L>\beta_g$, the eigenvalues can be either real or complex, suggesting that the eigenvalues corresponding to localized states may similarly exhibit real or complex characteristics.

Under OBCs, in Fig.~\ref {f6}(a2) with $a=0.5$, compared to Fig.~\ref{f6}(a1), a significant change in the fractal dimension $D_ {2}$ is observed when $n/L<\beta_ {g} $, indicating an alteration in the localization of the eigenstates. Figs.~\ref{f6}(a3) and \ref{f6}(a4) demonstrate that eigenstates with indices $n/L<\beta_ {g} $ transform SFL states, whereas those with $n/L>\beta_ {g} $ remain multifractal.
In Fig.~\ref{f6}(b2) with $a=2.0$, compared to Fig.~\ref{f6}(b1), the fractal dimension $D_ {2}$ undergoes a significant change for $n/L<\beta_{g}$, reflecting changes in the localization of the eigenstates. Figs.~\ref{f6}(b3) and \ref{f6}(b4) reveal that the eigenstates with indices $n/L<\beta_{g}$ become SFL states, while those with $n/L>\beta_{g} $ remain localized. These findings are consistent with those reported in reference \cite{PhysRevA.110.012222}. Moreover, comparing Fig.~\ref{f6}(a2) with Fig.~\ref{f6}(b2), a significant difference in fractal dimension $D_ {2}$ is evident for $n/L<\beta_{g} $, indicating variations in the degree of localization of the SFL state, which differs from the observation under PBCs. Furthermore, by comparing Figs.~\ref{f6}(a3) and \ref{f6}(b3), it is apparent that the degree of localization of the SFL state differs between long-range $(a<1)$ and short-range $(a>1)$ hoppings.
\section{ Results at larger sizes in the $P_2$ regime}\label{B}

 \begin{figure}[htp]
		 \centering
		\includegraphics[width=0.5\textwidth]{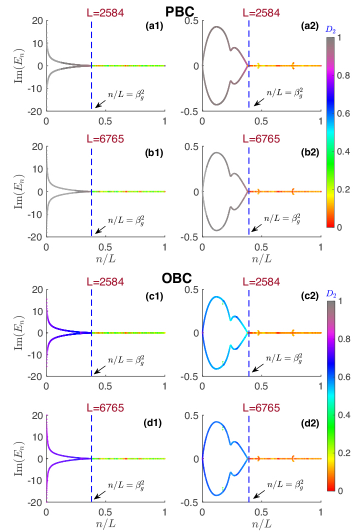}
		 \renewcommand\figurename{Fig}
		\renewcommand\figurename{Fig}
	\caption{ Imaginary part of the energy spectra, Im $(E_n )$,  under (a1)-(b1) PBCs and (c1)-(d1) OBCs for $a=0.5$ and $\Delta/J=2.6 $, with $L=2584$ and $L=6765$, respectively, sorted in ascending order of the real part. Similarly, the imaginary part of the energy spectra, Im $(E_n )$, under (a2)-(b2) PBCs and (c2)-(d2) OBCs for $a=2.0$ and $\Delta/J=3.6 $, with $L=2584$ and $L=6765$, respectively, also sorted in ascending order of the real part. The colorbar represents the fractal dimension $D_{2}$. For $L=2584$, we use $\beta_{g}=1597/2584 $ and $d=4$. For $L=6765$, we use $\beta_{g}=4181/6765 $ and $d=5$. }\label{f7}
	\end{figure}
    
For larger system sizes and other same parameters, we find that both in the PBCs and in OBCs are consistent with the phenomena described in the text.
Under PBCs, the eigenvalues for ergodic states are complex, while those for the multifractal states are real [Fig.~\ref {f7}(a1) and \ref {f7}(b1)],and the eigenvalues corresponding to localized states may similarly exhibit real or complex characteristics [Fig.~\ref {f7}(a2) and \ref {f7}(b2)]. Under OBCs,  compared to PBCs, the fractal dimension $D_ {2}$ undergoes a significant change for $n/L<\beta_{g}^2 $ in the case of  both $a<1$ and $a>1$, reflecting changes in the localization of the eigenstates. Moreover, comparing Fig.~\ref{f7}(c1) and \ref {f7}(d1) with Fig.~\ref{f7}(c2) and \ref {f7}(d2), a significant difference in fractal dimension $D_ {2}$ is evident for $n/L<\beta_{g}^2 $, indicating variations in the degree of localization of the SFL state, which differs from the observation under PBCs. At the same time, compared to the case of PBCs,  there was no significant change in the shape of the energy spectra, only in the magnitude of the values.

\section{Impact of long-range hopping on skin effect at larger sizes}\label{C}

 \begin{figure}[htp]
		
		\centering
		\includegraphics[width=0.5\textwidth]{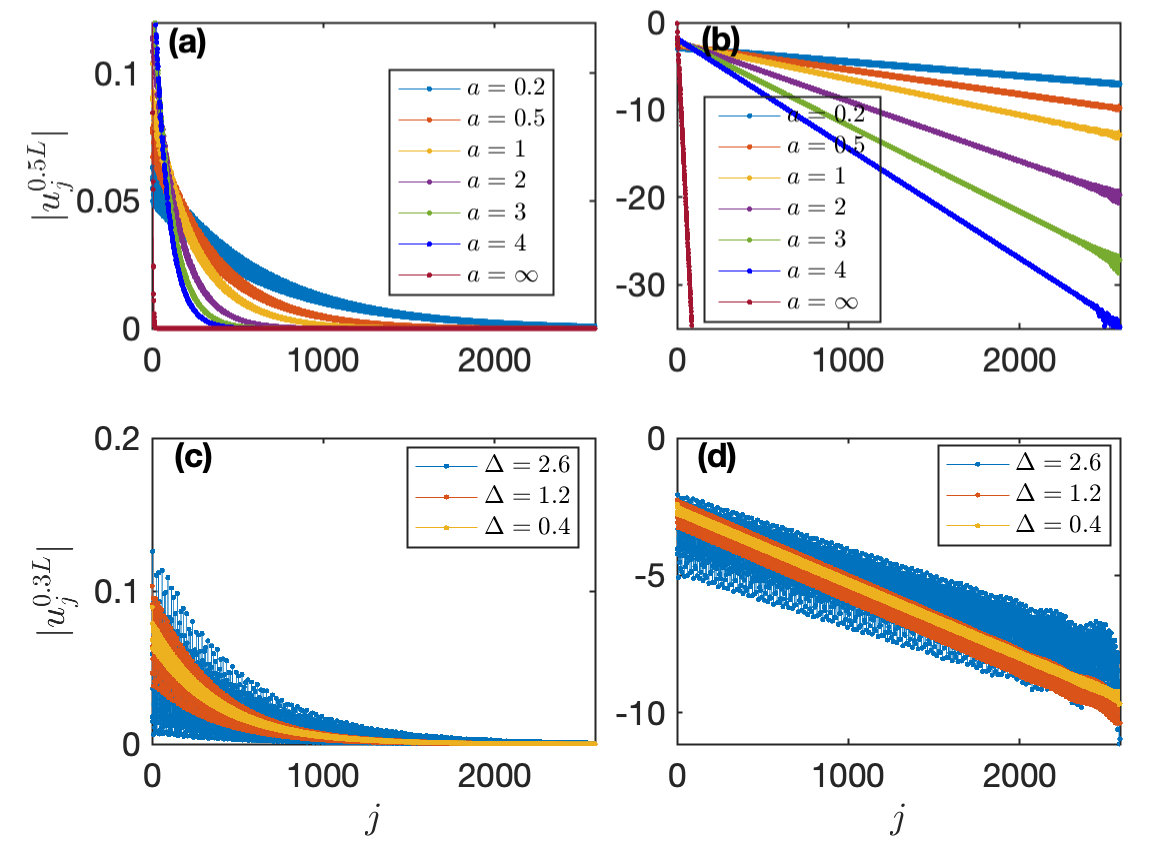}
		\renewcommand\figurename{Fig}
		
		 \renewcommand\figurename{Fig}
		 \caption{(a) Probability amplitude of the eigenstate corresponding to the $0.5L$$th$ eigenvalue and (b) its distribution on a logarithmic scale for different power-law indices $a$, with  parameters $\Delta/J=0.4$, $k=0.5$, $\beta _g = 1597/2584$,  and $L=2584$. (c) Probability amplitude of the eigenstate corresponding to the $0.3L$$th$ eigenvalue and (d) its logarithmic distribution for different strengths of the potential $\Delta $, with parameters $a=0.5$,  $k=0.5$,  $\beta _g = 1597/2584$,  and $L=2584$.}\label{f8}
	 \end{figure}
For larger system sizes and other same parameters, we find that impact of long-range hopping on skin effect are consistent with described in the text.
As shown in Fig.~\ref{f8}(a), for both $a>1$ and $a<1$, the states exhibit characteristics of SFL states. With an increase in the power-law index $a$, the manifestation of the skin effect becomes more pronounced, leading to a heightened degree of localization in the SFL state. At $a=\infty$, the eigenstate becomes fully localized at the left boundary.
 To more clearly illustrate the varying degrees of localization of SFL states for different $a$, we plot the logarithm of the probability amplitude of the $0.5L$$th$ eigenstate. As shown in Fig.~\ref{f8}(b), the probability amplitude decays more rapidly with increasing $a$. This  illustrates the influence of power-law index $a$ on the skin effect, demonstrating that in this system, long-range hoppings weaken the skin effect.
By fixing the power-law index $a=0.5 $, system size $L=2584$, and the golden mean $\beta_{g}=1597/2584$, we analyze probability amplitude distribution of the $0.3L$$th$ eigenstate and its logarithmic distribution across different  $\Delta $ values. As shown in Figs.~\ref{f8}(c) and \ref{f8}(d), variations in the potential strength do not significantly alter the degree of localization of the SFL state, even when viewed on a logarithmic scale. This suggests that the strength of the potential $\Delta $ does not affect the non-Hermitian skin effect.
\newpage
	\bibliography{re}

\end{document}